\documentclass[11pt,a4paper,onecolumn]{article}

\usepackage[T1]{fontenc}
\usepackage[utf8]{inputenc}
\usepackage{lmodern}
\usepackage[hmargin=28mm,vmargin=28mm]{geometry}
\usepackage{microtype}
\usepackage{amsmath}
\usepackage{amssymb}
\usepackage{gensymb}
\usepackage{booktabs} 
\usepackage{tikz}

\usepackage{graphicx}
\usepackage{subcaption}
\usepackage{siunitx}
\usepackage[round,sort]{natbib}
\usepackage{placeins} 

\usepackage{xcolor}
\definecolor{linkblue}{rgb}{0.10,0.20,0.50}
\usepackage[colorlinks=true,linkcolor=linkblue,citecolor=linkblue,urlcolor=linkblue,
            pdftitle={How problem formulation shapes simulation-based optimisation of tracked vehicle suspension},
            pdfauthor={Markus Pogulis, Martin Servin}]{hyperref}
\usepackage{doi}

\newcommand{\bg}[1]{\boldsymbol#1}

\makeatletter
\renewcommand{\@maketitle}{%
  \newpage\null\vskip -1.5em
  \begin{center}
    {\LARGE\bfseries \@title \par}\vskip 1.0em
    {\large \@author \par}
  \end{center}\par\vskip 0.8em}
\makeatother

\begin{document}

\title{How problem formulation shapes simulation-based optimisation of tracked vehicle suspension}

\author{%
  Markus Pogulis\textsuperscript{1,2}\qquad Martin Servin\textsuperscript{1,3}\\[0.6em]
  {\normalsize\normalfont
  \textsuperscript{1}Ume\aa\ University, SE-90187 Ume\aa, Sweden\\
  \textsuperscript{2}BAE Systems H\"agglunds AB, SE-89141 \"Ornsk\"oldsvik, Sweden\\
  \textsuperscript{3}Algoryx Simulation AB, SE-90736 Ume\aa, Sweden\\[0.4em]
  \href{mailto:markus.pogulis@umu.se}{markus.pogulis@umu.se},
  \href{mailto:martin.servin@umu.se}{martin.servin@umu.se}}}
\date{}

\maketitle

\begin{abstract}
Simulation-based optimisation is an established tool for vehicle suspension design, but how the optimisation problem should be formulated has received little attention. This paper investigates how the choice of objectives, evaluation scenarios and constraints influences the resulting design. A multibody model of a tracked vehicle with parameterised torsion-rod suspension and semi-active damping is combined with Bayesian optimisation to evaluate designs over representative terrain. Three formulations are investigated, progressing from single-objective ride-quality optimisation to a constrained multi-objective formulation balancing rough-terrain speed against pitch comfort on easy roads. The two single-objective formulations give similar designs that neglect comfort in routine driving, while the multi-objective formulation spans a trade-off from a middle-weighted, lightly damped comfort design to a forward-supported, more strongly damped performance design. The results demonstrate that problem formulation is a key component of simulation-based suspension design and should be considered as carefully as the suspension model itself.
\end{abstract}


\section{Introduction}
The suspension system largely determines the ride characteristics and off-road mobility of tracked vehicles operating over rough terrain. Its design directly influences the vibration transmitted to the crew, the dynamic loads acting on the vehicle structure, and ultimately the maximum speed at which the vehicle can safely and effectively traverse uneven ground. Tracked military and industrial vehicles are expected to operate over a broad spectrum of terrain conditions, ranging from relatively smooth roads to highly irregular off-road environments. Consequently, no single suspension design is optimal across all operating conditions.

Since evaluating suspension performance experimentally across a large range of terrain conditions is both time-consuming and expensive, multibody dynamics simulation has become an indispensable tool for suspension design. Different suspension designs can be evaluated under controlled and repeatable operating conditions, enabling systematic studies that would be impractical using physical prototypes alone. The predictive capability of high-fidelity multibody models has been demonstrated by \citet{rubinstein2004detailed}, who developed and experimentally validated a detailed tracked-vehicle model including individual track links, suspension components, and terrain interaction. This capability has shifted suspension design from manual parameter tuning towards systematic simulation-based optimisation, where candidate designs are evaluated automatically over a wide range of operating conditions.

Previous research has primarily focused on improving suspension concepts, understanding suspension dynamics, or optimising specific suspension parameters. \citet{gagneza2019estimation} simulated and measured the ride of a tracked combat vehicle with torsion-bar suspension and found that the torsion bars isolate most of the low-frequency vibrations. \citet{ata2014investigation} investigated the influence of damper placement during traversal of discrete terrain obstacles and showed that concentrating dampers at the vehicle extremities improves ride quality by reducing pitch-induced accelerations. \citet{thoresson2009part1,thoresson2009part2} optimised the spring and damper characteristics of an off-road vehicle modelled in MSC.ADAMS for ride comfort and handling, using simplified vehicle models to reduce the cost and numerical noise of the gradient computations. \citet{le2025vibration} used a dynamic interaction model of a tracked vehicle, its seat and its driver to show how the road excitation affects the driver's sighting efficiency and driving comfort. Both deteriorated with road roughness and speed, and they improved both by tuning the seat suspension with a multi-objective genetic algorithm. 

In these and related studies, ride quality is typically quantified using metrics such as RMS acceleration, vibration dose value (VDV), average absorbed power (AAP), or other measures derived from ISO 2631-1~\citep{ISO2631} and military ride evaluation procedures~\citep{TOP014B}. Accordingly, simulation-based suspension optimisation has generally been formulated by optimising one or more of these metrics over a predefined set of operating conditions. \citet{uys2007suspension} further investigated how the optimal suspension settings depend on where the ride comfort (measured as the weighted RMS vertical acceleration) is evaluated: at the driver, the rear passenger or both combined. They also varied the road and speed used for optimisation, demonstrating that these choices influence both the optimal design and its performance across operating conditions.

Beyond such comparisons, the formulation of the optimisation problem has received comparatively little systematic attention. Choices of ride metrics, terrain conditions, aggregation across operating conditions, and engineering constraints fundamentally determine which suspension characteristics are rewarded by the optimiser and therefore have a direct influence on the resulting design. Yet, in most published studies, these choices are treated as part of the experimental setup rather than as an object of scientific investigation. Instead, the emphasis has been on improving vehicle models, suspension concepts, or optimisation algorithms.

The objective of this paper is therefore not to propose another suspension model or optimisation algorithm, but to investigate how the formulation of a simulation-based optimisation problem influences the resulting suspension design. Using computationally expensive multibody dynamics simulations together with multi-objective Bayesian optimisation, we successively develop three optimisation formulations that differ in their objective functions, operating scenarios and engineering constraints. This extends the comparison of \citet{uys2007suspension} to different types of objectives, to many roads and speeds combined in one objective, and to constraints on the static attitude. We show that the two single-objective formulations, one minimising vibration and one maximising rough-terrain speed, lead to similar designs. Adding a comfort objective on easy roads instead leads to substantially different designs and an explicit trade-off, while the vehicle model and optimisation algorithm remain the same throughout. Furthermore, analysis of the Pareto-optimal designs provides physical insight into how the torsion-rod layout, damping characteristics, and vehicle static attitude govern the trade-off between pitch comfort and rough-terrain speed, and how ride quality constraints limit the comfort-oriented designs.

\section{Vehicle model and simulation}
\label{sec:modeling}
This section introduces the computational model used throughout the optimisation study. It first describes the tracked-vehicle model, with emphasis on the suspension system and simplified track representation, before presenting the numerical multibody simulation method.

\subsection{Vehicle model}
\begin{figure}[tbp]
  \centering
  \includegraphics[width=0.8\textwidth]{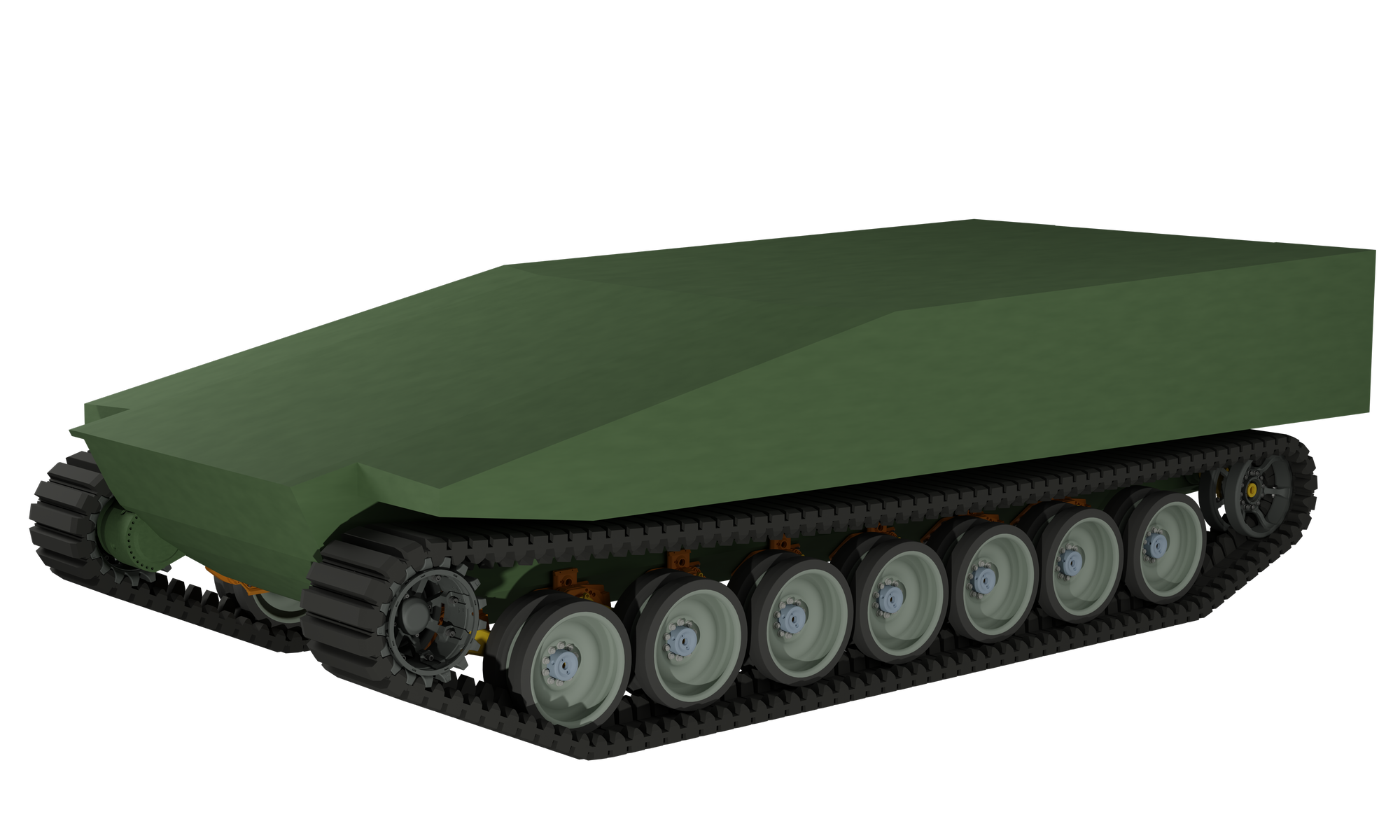}
  \caption{Illustration of the full tracked-vehicle model.}
  \label{fig:vehicle}
\end{figure}

The vehicle, illustrated in Fig.~\ref{fig:vehicle}, is represented as a rigid multibody system comprising the chassis, swing arms, road wheels, drive sprocket, idler, suspension system, and track.
Since the objective of this study is suspension optimisation on rigid terrain, the suspension system is modelled in detail, whereas the track is represented using a computationally efficient simplified model.
The total vehicle mass is approximately 33 tonnes, of which the chassis accounts for about 30 tonnes. The vehicle is 6.0 m long, 2.8 m wide, and 1.7 m high.

\subsubsection{Suspension system}
\label{sec:suspension}
\begin{figure}[tbp]
  \begin{subfigure}[t]{0.47\textwidth}
    \centering
    \begin{tikzpicture}
      \node[anchor=south west, inner sep=0] (img) at (0,0)
        {\includegraphics[width=0.98\linewidth]{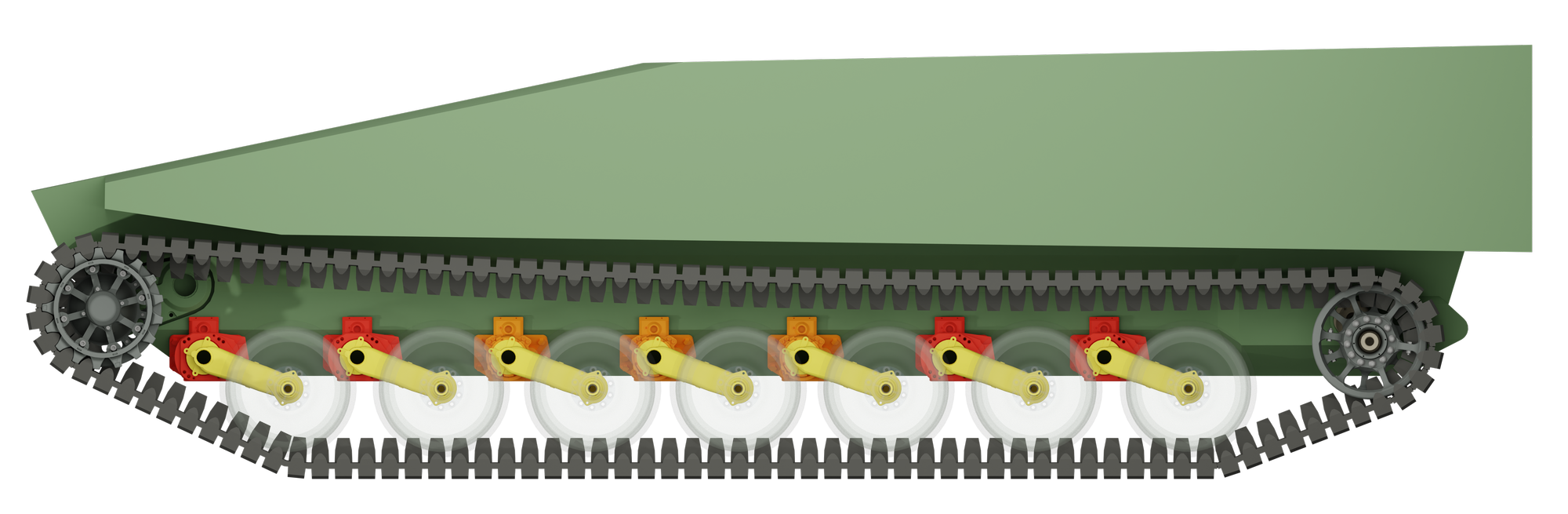}};
      \begin{scope}[x={(img.south east)}, y={(img.north west)}]
        \coordinate (driver) at (0.26,0.44);
        \filldraw[fill=white, draw=black, line width=0.8pt]
            (driver) circle (3pt);
        \draw[black, line width=0.8pt]
            ([xshift=-2pt,yshift=-2pt]driver) --
            ([xshift= 2pt,yshift= 2pt]driver)
            ([xshift=-2pt,yshift= 2pt]driver) --
            ([xshift= 2pt,yshift=-2pt]driver);
      \end{scope}
    \end{tikzpicture}
  \end{subfigure}
  \begin{subfigure}[t]{0.47\textwidth}
    \centering
    \begin{tikzpicture}
      \node[anchor=south west, inner sep=0] (img) at (0,0)
        {\includegraphics[width=0.98\linewidth]{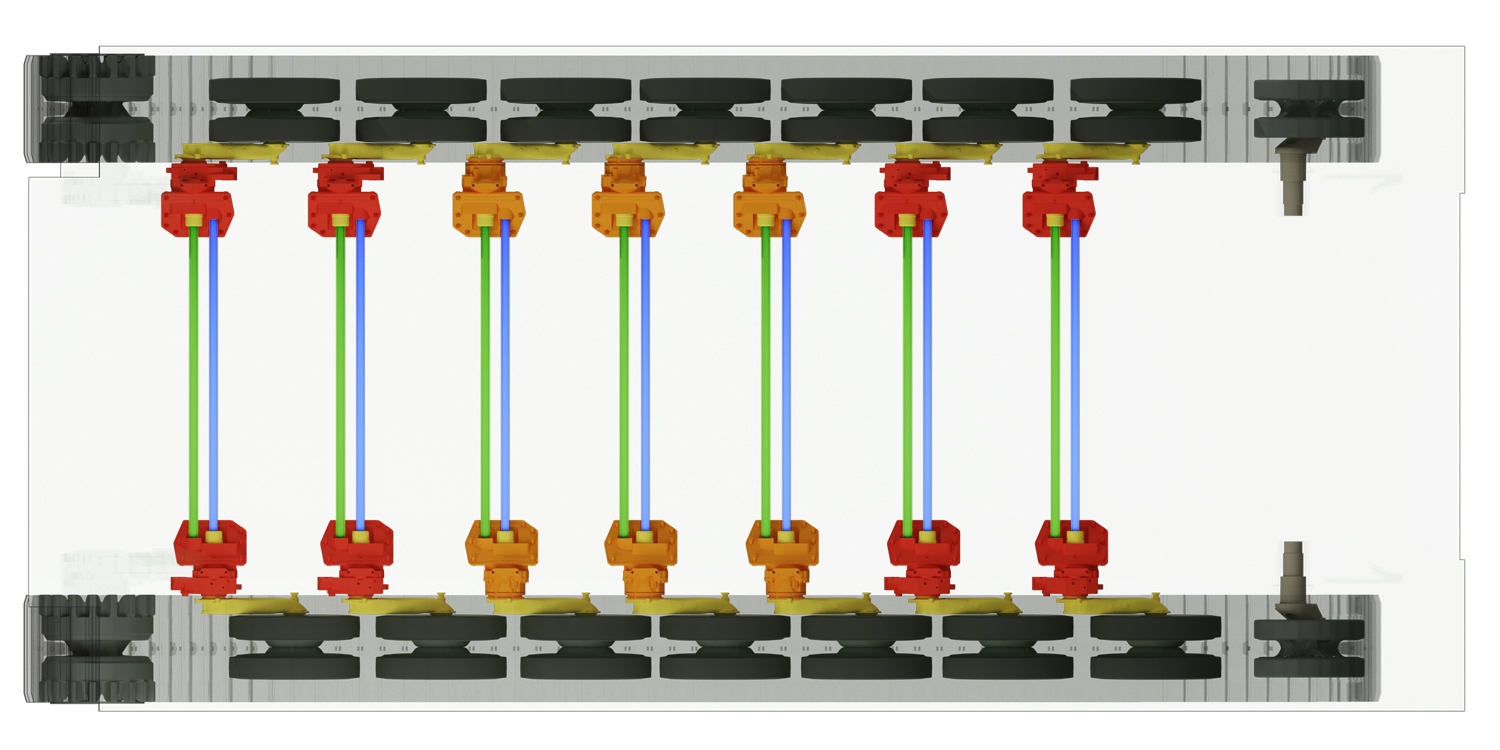}};

      \begin{scope}[x={(img.south east)}, y={(img.north west)}]
        \coordinate (driver) at (0.26,0.35);
        \filldraw[fill=white, draw=black, line width=0.8pt]
            (driver) circle (3pt);
        \draw[black, line width=0.8pt]
            ([xshift=-2pt,yshift=-2pt]driver) --
            ([xshift= 2pt,yshift= 2pt]driver)
            ([xshift=-2pt,yshift= 2pt]driver) --
            ([xshift= 2pt,yshift=-2pt]driver);
      \end{scope}
    \end{tikzpicture}
  \end{subfigure}\\

  \caption{Side view and top view showing the seven swing arms (yellow) supported by a torsion rod, where the green rods are connected to the right-hand swing arms and the blue rods to the left-hand ones. The two frontmost and two rearmost stations (red) are semi-actively damped.
  The driver position is marked by $\otimes$.}
  \label{fig:vehicle_to_side}
\end{figure}

Each side of the vehicle is supported by seven road-wheel stations, each comprising a road wheel mounted on a trailing swing arm (Fig.~\ref{fig:vehicle_to_side}). The stations are numbered from 1 to 7 in the forward driving direction, with station 1 at the front of the vehicle.

\begin{figure}[tbp]
    \centering
    \includegraphics[width=0.6\textwidth]{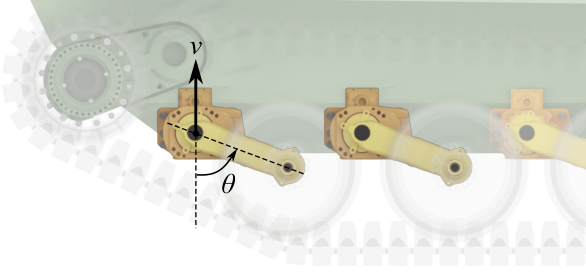}
    \caption {Close-up illustration of the front suspension station with trailing-arm rotation angle $\theta$ and vertical station velocity $v$.}
    \label{fig:damping_schematic_view}
\end{figure}

\paragraph{Torsion-rod suspension}
Each road-wheel station is supported by a torsion-rod suspension acting about its trailing-arm pivot. At the front and rear stations, 1--2 and 6--7, a semi-active frictional damper acts in parallel with the torsion rod, while the three middle stations 3--5 are undamped. The total station torque is thus the sum of the torsion-rod and damping torques,
\begin{equation}
    T(\theta,v) = T_s(\theta) + T_d(v),
    \label{eq:total_torque}
\end{equation}
where \(\theta\) is the trailing-arm rotation, positive in compression, as marked in Fig.~\ref{fig:damping_schematic_view}.
The torsion rod is modelled as linear elastic torsional spring,
\begin{equation}
    T_s(\theta) = k_t \left(\theta - \theta_0\right),
    \label{eq:torsion_torque}
\end{equation}
with stiffness $k_t = GJ/L = G\pi d^4/(32L)$ where \(G\) is the shear modulus, \(L\) is the effective rod length, \(d\) is the rod diameter, and \(J=\pi d^4/32\) is the polar moment for a solid circular shaft. 
A maximum allowable shear stress \(\tau_{\max}\) limits the twist to $\theta_{\max} = 2\tau_{\max}L/(Gd)$.
The installation angle \(\theta_0\) is selected so that the maximum twist is reached at the bump stop angle \(\theta_{\text{bump}}\), the trailing-arm angle at full compression. In this work \(\theta_{\text{bump}} = 105\degree\), which places the arm 15\degree\ above horizontal. The installation angle becomes $\theta_0 = \theta_{\text{bump}} - \theta_{\max}$ which gives a pre-load at static equilibrium. The angle \(\theta_0\) is therefore fixed and is not treated as a design variable for optimisation.

\begin{figure}[tbp]
    \centering
    \includegraphics{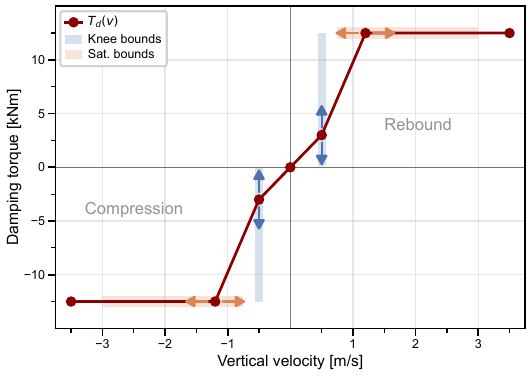}
    \caption{Damping torque $T_d(v)$ as a function of vertical station velocity $v$ for the semi-active damping, with compression at negative $v$ and rebound at positive $v$. The shaded bands and arrows mark the search region for the compression and rebound knee torques (blue), $T_\text{c}$ and $T_\text{r}$, and compression and rebound saturation velocities (orange), $v_\text{c}$ and $v_\text{r}$.}
    \label{fig:baseline_damping}
\end{figure}

\paragraph{Semi-active frictional damping}
The damper produces a torque $T_d(v)$ that depends on the vertical station velocity $v$ illustrated in Fig.~\ref{fig:baseline_damping}. 
Being frictional, it is purely dissipative: it cannot drive the swing arm, only resist its motion.  The compression stroke (negative $v$, lower left) and the rebound stroke (positive $v$, upper right) are shaped independently. 
The damping characteristic curve has four parameters, the compression and rebound knee torques, $T_\text{c}$ and $T_\text{r}$, at vertical station velocity $\pm 0.5$ m/s, and the compression and rebound saturation velocities, $v_\text{c}$  and $v_\text{r}$, where the torque saturates to $\pm 12.5$ kNm. 
To represent the response time of a real damper, $v$ is passed through a first-order low-pass filter before $T_d(v)$ is evaluated, giving the torque a short lag of about 100~ms.

\subsubsection{Simplified track model}
\begin{figure}[tbp]
  \centering
  \includegraphics[width=0.8\textwidth]{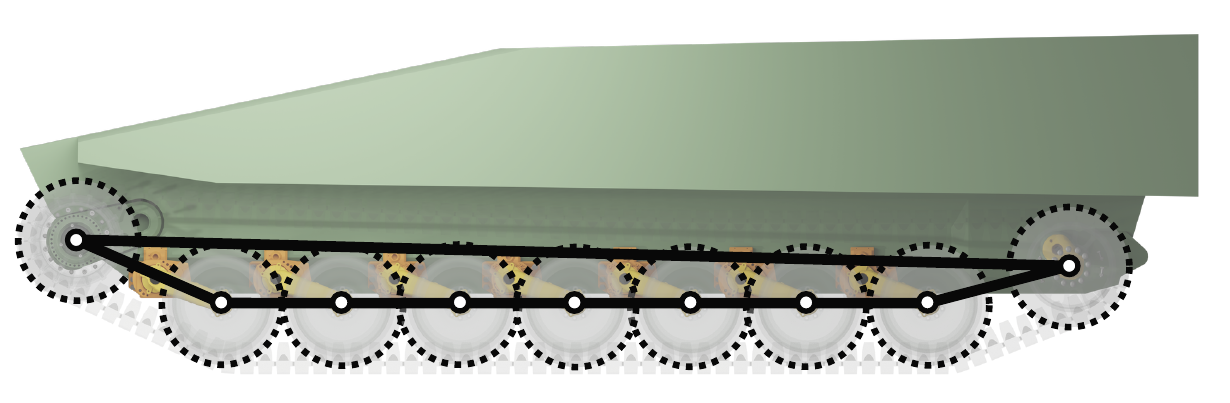}
  \caption{Illustration of the simplified wire-track model. The road wheels are suspended by the swing arms and interconnected by a tensioned wire forming a closed loop with the sprocket and idler.}
  \label{fig:tracks}
\end{figure}

To reduce computational cost, the track is represented by a simplified wire-track model, illustrated in Fig.~\ref{fig:tracks}, rather than as a chain of discrete links. The wire is a massless elastic string~\citep{servin2007massless}, starting at the sprocket and threaded through frictionless routing nodes at the wheel centres of the seven road wheels and the idler before returning to the sprocket, forming a closed, tensioned loop. The wire has an elastic stiffness of 1000 kN/m and its tension is initialised to 25 kN by re-adjusting the swing arm holding the idler wheel. The thickness of the real track is accounted for by an equivalent increase in wheel radius. The mass of the track is redistributed to the nearest running-gear bodies, with most attributed to the sprocket and idler, which carry the overhang. All road wheels have the same rotational speed as the drive sprocket.

The wire-track model belongs to the same family of continuous-loop representations as the string track model of MSC Adams Tracked Vehicle \citep[ATV;][]{adamsATV}. This idealisation is justified by the scope of the study. The wheel loads are carried by the suspension, which is also the main subject of the optimisation, while on the rigid ground and relatively smooth terrain considered here, the track plays only a secondary role in ride quality. The swing-arm motion that governs the ride metrics is therefore expected to be only weakly affected by the track idealisation.
The alternative, resolving the tracks as a multi-link model, would approximately add 250 rigid bodies and constraints and require a smaller integration time-step to resolve their higher-frequency dynamics and ground contacts. 

\subsection{Simulation method}
The vehicle dynamics is simulated using AGX Dynamics~\citep{AGX2026}, a commercial simulation software for multibody and multiphysical systems with nonsmooth dynamics in maximal (descriptor) coordinates. Model parameters include geometric shape and inertial parameters of each body, joint kinematics, powertrain characteristics, and contact material properties, including elasticity, friction, and restitution. The terrain is represented by a static height field with a regular grid resolution of 0.1 m.

The constrained equations of motion are integrated using a fixed time step of 0.001 s and using AGX's hybrid direct–iterative solver.

\section{Performance evaluation}
\label{sec:performance-evaluation}
To evaluate suspension performance consistently across different operating conditions, representative road profiles and quantitative ride-quality metrics are required. The road profiles define the operating scenarios used throughout the study, while the ride-quality metrics quantify the resulting vehicle response. These evaluation components form the basis for the optimisation problem.

\subsection{Road profiles}
\label{sec:roads}

A vehicle's ride quality depends on how its suspension responds to the road profiles it drives over. Evaluating the suspension therefore calls for a range of roads that excite it in different ways, rather than a single hand-picked profile.

In this work we use a set of twelve road profiles, five measured and seven generated, organised into two sets. The rough set represents more demanding terrain and off-road driving, and the easy set represents milder terrain and routine driving. Each road is described by two quantities, a roughness amplitude given by the root-mean-square of the detrended profile~\citep{bylsma2014spacial} and a spectral shape following ISO~8608~\citep{ISO8608}. Each road is laterally uniform, with the same height profile for the left and right track, and therefore excites heave and pitch but no direct roll. We first present the measured roads and their characteristics, then the generated rough and easy variants.

\subsubsection{Measured roads}
\label{sec:measured-roads}
The road set is built on five road profiles from various proving grounds, spanning a range of terrain severities. They were sampled at a spacing $\Delta x_{\mathrm{src}}$ of about 0.3~m and range in length $L$ from 91 to 205~m. Each road is labelled by a subscript equal to its $\mathrm{RMS}_r$ in millimetres to one decimal, prefixed by $R$ (rough set) or $E$ (easy set), so the five measured roads are $R_{29.8}$ through $R_{75.6}$.

Table~\ref{tab:original_roads} lists the five roads with their characteristics. The detrended RMS ranges from 30 to 76~mm and the waviness exponent $w$ from 2.10 to 2.96, within the 2 to 4 range historically observed for ride-quality courses in TOP~01-1-010A~\citep{TOP010A}. These characteristics are defined in the following section.

\begin{table}[tbp]
\centering
\small
\setlength{\tabcolsep}{4pt}
\caption{Characteristics of the original measured roads.}
\label{tab:original_roads}
\begin{tabular}{lccccc}
\toprule
ID & $\Delta x_{\mathrm{src}}$ (m) & $L$ (m) & $\mathrm{RMS}_r$ (mm) & $w$ & Class \\
\midrule
$R_{29.8}$ & 0.305 & 121.9 & 29.8 & 2.96 & F \\
$R_{33.3}$ & 0.305 & 103.9 & 33.3 & 2.10 & E \\
$R_{38.0}$ & 0.305 & 152.4 & 38.0 & 2.74 & F \\
$R_{52.7}$ & 0.305 &  91.4 & 52.7 & 2.55 & F \\
$R_{75.6}$ & 0.200 & 205.0 & 75.6 & 2.78 & G \\
\bottomrule
\end{tabular}
\end{table}

\subsubsection{Road profile characteristics}
\label{sec:road-chacteristics}
Each measured road profile $F(x)$ is split into a long-wavelength trend $\bar F(x)$ and a short-wavelength roughness $F_r(x)$,
\begin{equation}
    F(x) = \bar F(x) + F_r(x),
    \label{eq:decomposition}
\end{equation}
where the trend is captured by a double-sided exponential moving average of the profile~\citep{bylsma2014spacial}. A characteristic wavelength sets the boundary between trend and roughness, and we use 3~m throughout. The detrended roughness profile is then computed as the residual $F_r = F(x) - \bar F(x)$. The roughness amplitude is the root-mean-square of $F_r$,
\begin{equation}
    \mathrm{RMS}_r = \sqrt{\frac{1}{N}\sum_{i=1}^{N} F_r(x_i)^2},
    \label{eq:rms_r}
\end{equation}
%
which is reported in Table~\ref{tab:original_roads} for the measured roads.

The spectral shape describes how the roughness is distributed over wavelengths. The displacement power spectral density $G_d(n)$ is estimated as a function of spatial wavenumber $n$ using Welch's method with the standard parameters of TOP~01-1-010A~\citep{TOP010A}, and is fit to the power-law form of ISO~8608~\citep{ISO8608},

\begin{equation}
    G_d(n) = G_d(n_0)\left(\frac{n}{n_0}\right)^{-w}, \qquad n_0 = 0.1~\mathrm{cyc/m}.
    \label{eq:iso8608}
\end{equation}
The unevenness index $G_d(n_0)$, evaluated at the 10~m reference wavelength, sets the ISO class A--H, and the waviness exponent $w$ describes the spectral slope.

\subsubsection{Generated roads}
\label{sec:generated-roads}

The seven generated roads are constructed from the measured ones. By scaling the trend and roughness components of Eq.~\eqref{eq:decomposition}, we generate two additional rough roads and the five roads of the easy set.

\paragraph{Generated rough variants}
To fill out the upper half of the roughness range, two additional rough roads are generated. Given a source profile $F = \bar F + F_r$ on $x \in [0, L_0]$, a generated rough variant is
\begin{equation}
F'(x') = a_r\,F_r(x) + \bar F(x), \qquad x' \in [0,\, \alpha L_0],
\label{eq:rough_gen}
\end{equation}
with roughness amplification $a_r = 1.8$ and length stretch $\alpha = 1.2$. The trend is preserved, and the length extension is implemented by uniform resampling, so the bumps of the source road keep their order but are spaced slightly further apart. The two variants are constructed from sources $R_{33.3}$ and $R_{38.0}$, and their characteristics are given in Table~\ref{tab:rough_variants}. Together with the five measured roads, they complete the rough set of seven roads, spanning an $\mathrm{RMS}_r$ range of 30 to 76~mm and waviness exponents of 2.1 to 3.0, all of ISO class E to G.

\begin{table}[tbp]
\centering
\small
\setlength{\tabcolsep}{4pt}
\caption{Characteristics of the generated rough variants, constructed from their source roads by amplifying $F_r$ by $a_r = 1.8$ and extending the length by $\alpha = 1.2$.}
\label{tab:rough_variants}
\begin{tabular}{lccccc}
\toprule
ID & Source & $L$ (m) & $\mathrm{RMS}_r$ (mm) & $w$ & Class \\
\midrule
$R_{52.5}$ & $R_{33.3}$ & 124.7 & 52.5 & 2.18 & F \\
$R_{61.8}$ & $R_{38.0}$ & 182.9 & 61.8 & 2.36 & F \\
\bottomrule
\end{tabular}
\end{table}

\paragraph{Generated easy variants}
The five easy roads are milder variants of the measured ones, with the roughness amplitude reduced and the trend almost entirely removed,
\begin{equation}
F''(x) = b_r\,F_r(x) + b_{\bar F}\,\bar F(x), \qquad b_{\bar F} = 0.05,
\label{eq:easy_gen}
\end{equation}
with the roughness scale $b_r$ chosen per source in 0.40 to 0.55 and the length preserved. Table~\ref{tab:easy_variants} reports the characteristics, with $\mathrm{RMS}_r$ spanning 12 to 22~mm.

Removing the trend leaves the roughness of the source road without its slower terrain undulations. Because the ISO~8608 fit anchors on this long-wavelength content, the easy roads show consistently lower waviness than their sources.

\begin{table}[tbp]
\centering
\small
\setlength{\tabcolsep}{4pt}
\caption{Characteristics of the generated easy variants, constructed from their source roads by scaling $F_r$ by $b_r$ and $\bar F$ by $b_{\bar F} = 0.05$, length unchanged.}
\label{tab:easy_variants}
\begin{tabular}{lccccccc}
\toprule
ID & Source & $b_r$ & $L$ (m) & $\mathrm{RMS}_r$ (mm) & $w$ & Class \\
\midrule
$E_{12.5}$ & $R_{29.8}$   & 0.50 & 121.9 & 12.5 & 2.43 & D \\
$E_{18.0}$ & $R_{33.3}$   & 0.55 & 103.9 & 18.0 & 1.71 & D \\
$E_{18.1}$ & $R_{38.0}$   & 0.50 & 152.4 & 18.1 & 2.27 & D \\
$E_{21.8}$ & $R_{52.7}$   & 0.50 &  91.4 & 21.8 & 2.01 & D \\
$E_{17.9}$ & $R_{75.6}$   & 0.40 & 205.0 & 17.9 & 2.17 & D \\
\bottomrule
\end{tabular}
\end{table}

\subsection{Ride quality metrics}
\label{sec:metrics}
This section summarises the metrics used to quantify ride quality in this work, the vibration dose value (VDV) of ISO~2631-1~\citep{ISO2631} and the average absorbed power \citep[AAP;][]{lee1968analytical,TOP014B}. We adopt the methodologies of these standards and use them to compare and rank suspension designs rather than to assess human exposure formally.

ISO~2631-1 defines whole-body exposure from a weighted combination of the translational and rotational axes, but for a wheeled military vehicle on off-road terrain, the vertical acceleration has been found to give the most reliable correlation with subjective ride comfort~\citep{SchalkEls2005}. The vertical metrics, VDV and AAP, are therefore evaluated at the driver position (Fig.~\ref{fig:vehicle_to_side}) in the vertical direction ($z$) only. Since the driver sits towards the front and left of the vehicle, this vertical signal also carries part of the chassis pitch and roll rather than pure heave.

\subsubsection{Vibration dose value and ISO2631-1}
\label{sec:vdv}

The vibration dose value is a cumulative measure of vibration exposure,
\begin{equation}
    \mathrm{VDV} = \left( \int_0^{T} a_w(t)^4 \, \mathrm{d}t \right)^{1/4},
    \label{eq:vdv}
\end{equation}
where $a_w(t)$ is the frequency-weighted acceleration and $T$ the exposure duration. For the driver vertical acceleration we apply the $W_k$ weighting of ISO~2631-1 before evaluating Eq.~\eqref{eq:vdv}, giving the driver vertical dose $\mathrm{VDV}_z$. The fourth power makes the dose sensitive to acceleration peaks, which suits off-road driving with its occasional large shocks. ISO~2631-1 itself recommends the dose method over the basic root-mean-square evaluation for motion with such peaks.

Because the dose accumulates over time, it grows with the exposure duration. The roads in our set differ in length, so at a given speed they correspond to different simulated durations. To compare designs on a common basis, each dose is normalised to a 30-minute reference duration by the factor $(T_{\mathrm{ref}}/T)^{1/4}$, where $T$ is the simulated duration and $T_{\mathrm{ref}} = 30$~min. All reported VDV values use this normalisation.

The same dose quantifies chassis pitch severity. The pitch (rotational) acceleration is frequency-weighted with the ISO~2631-1 $W_e$ weighting, then evaluated and normalised identically, giving the pitch dose $\mathrm{VDV}_{\mathrm{pitch}}$.

For context, ISO~2631-1 places its health guidance caution zone at VDV values of roughly 8.5 to 17~m/s$^{1.75}$ over an 8-hour exposure. These limits were not developed for heavy vehicles on rough terrain. For a wheeled military vehicle, \citet{SchalkEls2005} reports a 50\% subjective-discomfort level near 26~m/s$^{1.75}$ over 4~hours, well above the ISO guideline. 

\subsubsection{Average absorbed power}
\label{sec:aap}

Average absorbed power (AAP) measures the rate at which vibration transfers mechanical energy to the human body. Where the VDV accumulates a dose and emphasises transients, the AAP captures the continuous workload the ride imposes on the occupant, and is a standard measure in military ride-dynamics evaluation~\citep{TOP014B,murphy1986comparison}.

Following TOP~01-1-014B~\citep{TOP014B}, the AAP is computed from the spectrum of the driver's vertical acceleration,
\begin{equation}
    \mathrm{AAP} = \sum_{i=1}^{n} C_i\, A_i^2,
    \label{eq:aap}
\end{equation}
where $A_i$ is the rms acceleration within the $i$th spectral band and $C_i$ a frequency-dependent absorbed-power coefficient that weights the spectrum towards the frequencies to which the human body is most sensitive~\citep{lee1968analytical}.

Ride limits on rough terrain are commonly expressed as a tolerance on the absorbed power at the driver's seat. TOP~01-1-014B~\citep{TOP014B} sets this tolerance at 6~W, a level that typically coincides with the onset of suspension bottoming. It is a practical operational limit rather than a human tolerance limit~\citep{murphy1986comparison}.

\section{Optimisation}
\label{sec:methodology}
This study employs simulation-based optimisation to identify suspension designs that improve ride performance over a range of operating conditions. This section defines the optimisation problem, including the design variables and simulation scenarios, before describing the Bayesian optimisation framework used to solve it.

\subsection{Problem formulation}
\label{sec:formulation}
The optimisation goal is to tune the suspension using expensive multibody simulations: each evaluation of a candidate design requires simulating the vehicle over many road--speed combinations. The main design choice in this project was therefore not only the parameterisation of the suspension, but also the objective definition itself, since what we choose to optimise strongly shapes the type of suspension the optimiser prefers.

\subsubsection{Scenarios and evaluation cohorts}
\label{sec:scenarios}

A single simulation is defined by a \emph{scenario} $s = (r,v)$ consisting of a road $r\in\mathcal{R}$ traversed in a given direction at a given speed $v\in\mathcal{V}$. Each road can be traversed in either direction, and the two directions excite the suspension differently, so they count as distinct scenarios. A candidate design is evaluated by aggregating ride metrics over a set of scenarios
\begin{equation}\label{eq:scenarios}
    \mathcal{S} = \left\{ (r,v) \mid r\in\mathcal{R}, v\in\mathcal{V}\right\}
\end{equation}
so that the optimisation cannot tune the suspension to a single hand-picked terrain/speed combination. The speed of a scenario is either prescribed $v\in\mathcal{V}(r)$ or, for the limiting-speed objective introduced below, given by the maximum speed not violating a given limit for the average absorbed power
\begin{equation}\label{eq:v_max}
    v_{\max}(r) = \max_{v\in \mathcal{V}_\text{search}}\left\{v \mid
    \mathrm{AAP}(r,v)\le \mathrm{AAP}_\text{max}\right\}.
\end{equation}
Such a limiting speed is a standard measure of vehicle mobility on rough terrain~\citep{murphy1986comparison}. The set of roads $\mathcal{R}$ consists of the rough roads $\mathcal{R}_\mathrm{rough}$ and easy roads $\mathcal{R}_\mathrm{easy}$ introduced in Section~\ref{sec:roads}. 

\subsubsection{Design variables}
\label{sec:design_vars}

The design vector $\bg{x}\in\mathcal{X}\subset\mathbb{R}^{11}$ collects the suspension-system parameters subject to optimisation. Seven variables are the torsion-rod diameters $d_1,\dots,d_7\in[35,55]$~mm of the road-wheel stations. The left and right rod of each station share the same diameter, so the 14 physical rods reduce to seven variables. 

The remaining four variables shape the semi-active damping curve of Fig.~\ref{fig:baseline_damping}, which maps the vertical station velocity to a damping torque and is shared by the damped stations.

\subsection{Objectives}
\label{sec:formulation:objectives}
This section summarises the objective formulations explored in the project and the motivation for moving from one to the next. The detailed outcome and design trends are presented later in Section~\ref{sec:results}; here we focus on what each objective incentivises and why it was refined.

\subsubsection{A: Vibration minimisation}
\label{sec:argminVDV}
Given ISO~2631-1, it is natural to minimise the vibration as measured by the VDV at the driver's location. Specifically, we choose as objective the driver's vertical VDV  (Section~\ref{sec:metrics})
\begin{equation}\label{eq:objective-vibration}
        \min_{\bg{x}} \;\; F_{\mathrm{VDV}}(\bg{x})
    = \sum_{(r,v)\in \mathcal{S}_{A}} w_r\,\mathrm{VDV}_{z}(r,v,\bg{x}),
\end{equation}
where the weight $w_r = \big(\mathrm{RMS}_r/\mathrm{RMS}_0\big)^2$ with $\mathrm{RMS}_0= 40$~mm puts larger emphasis on roads with larger roughness. The set of scenarios $\mathcal{S}_A$ combines the seven rough roads, traversed in both directions, with the five easy roads. Each road is traversed at different speeds $v\in\mathcal{V}(r)$.

This objective is straightforward to implement and directly targets a standard dose-type severity measure. However, there are two
issues with this objective:
\textit{i) Manual speed selection:} The objective requires specifying road--speed combinations $\mathcal{V}(r)$ a priori. If some speeds are unrealistically high, the objective might be dominated by scenarios that are effectively infeasible to ``fix'' by tuning the suspension system alone.
\textit{ii) Peak sensitivity and bias:} VDV strongly penalises short high-acceleration events. In off-road simulation this can over-emphasise rare shocks (e.g., hard contacts) relative to sustained ride quality. This is not ``wrong'', but it is a specific design bias that can dominate the optimisation if not controlled.

These observations led us to investigate an alternative, more mission-oriented, objective: rather than minimising severity at preselected speeds, we search for the highest speed that stays within a tolerance.

\subsubsection{B: Limiting speed maximisation}
\label{sec:argmaxV}
To eliminate reliance on preselected road-speed pairs, we examined the objective of maximum AAP limiting speed 
\begin{equation}\label{eq:objective-speed}
    \max_{\bg{x}} \;\; F_{\mathrm{speed}}(\bg{x}) =
    \sum_{r\in \mathcal R_{\mathrm{rough}}} w_r \, v_{\max}(r,\bg{x}),
\end{equation}
where $v_{\max}(r,\bg{x})$ is found for each road $r$ and design $\bg{x}$ by a search over a set of speeds, in which each evaluation of $\mathrm{AAP}(r,v,\bg{x})$ requires a simulation of the road at speed $v$. The weights $w_r$ are the same as in formulation A, and the objective is evaluated over the rough roads only, where $v_{\max}$ can vary more. This formulation better matches the operational question ``how fast can we go on rough terrain?'' and eliminates the sensitivity to how individual road--speed points are chosen. However, the objective emphasises speed as performance and does not explicitly encode comfort on easier terrain. This motivated the introduction of a separate comfort objective.

\subsubsection{C: Speed and comfort}
\label{sec:MOBO}
Ride quality is a trade-off problem: improved rough-road speed capability can come at the cost of comfort in more routine driving. We therefore extend the speed objective of B with a comfort objective and pose the multi-objective problem
\begin{align}
        \max_{\bg{x}\in\mathcal{X}} \;\;& F_{\mathrm{speed}}(\bg{x})
          = \sum_{r\in \mathcal{R}_{\mathrm{rough}}} w_r \, v_{\max}(r,\bg{x}), \label{eq:mobo-speed}\\
        \min_{\bg{x}\in\mathcal{X}} \;\;& F_{\mathrm{comf}}(\bg{x})
          = \sum_{(r,v)\in \mathcal{S}_{\mathrm{easy}}} \mathrm{VDV}_{\mathrm{pitch}}(r,v,\bg{x}),
    \label{eq:mobo-comf}
\end{align}
where the comfort objective sums the pitch VDV, computed from the chassis pitch motion (Section~\ref{sec:metrics}), over the scenarios $\mathcal{S}_{\mathrm{easy}}$ consisting of the easy roads traversed with speeds $\mathcal{V}_\mathrm{easy}$. These scenarios are intended to represent routine driving rather than mobility limits. 

Solving Eqs.~\eqref{eq:mobo-speed}--\eqref{eq:mobo-comf} yields not a single optimum but a set of Pareto-optimal designs, each representing a different balance between the two objectives.

\subsubsection{Ride quality constraints}
\label{sec:constraints}
The ride metrics are indifferent to the static attitude and ride height of the vehicle.  We found that unconstrained optimisation often traded both for ride-metric improvements. Therefore, we introduced constraints on the vehicle's static equilibrium configuration on flat ground 
\begin{align}\label{eq:constraint-pitch}
    |\phi(\bg{x})| & \le \phi_\mathrm{max}, \\      
    \label{eq:constraint-clearance}
    c(\bg{x}) & \ge c_\mathrm{min},      
\end{align}
where $\phi(\bg{x})$ is the vehicle pitch angle and $c(\bg{x})$ is the minimum ground clearance along the vehicle chassis.

\subsection{Bayesian optimisation}
Evaluating a candidate suspension design is computationally expensive, since each evaluation requires a large number of multibody simulations over different road and speed combinations. We therefore employ Bayesian optimisation (BO), a sample-efficient method for expensive black-box problems. BO fits probabilistic surrogate models of the objective and constraint values to previously evaluated designs and uses an acquisition function to select new designs that balance exploration and exploitation.

The optimisation is implemented in the Ax framework~\citep{olson2025ax} and coupled to AGX Dynamics through a Python interface. It starts from initial designs drawn from a Sobol sequence, 50 for the single-objective formulations A and B and 60 for the multi-objective formulation C. Each objective and constraint is then modelled by a Gaussian process with the sparse axis-aligned subspace (SAAS) prior~\citep{eriksson2021saas}, and new designs are proposed one at a time using the qLogNEI acquisition function for A and B and qLogNEHVI for C~\citep{balandat2020botorch,ament2023unexpected,daulton2021parallel}. Each proposed design is simulated over the scenarios of its formulation, and the resulting objective and constraint values are returned to Ax, which updates the surrogate models before proposing the next design. In total, 200 designs were evaluated for each of A and B, and 455 for C.


\section{Simulation setup}
\label{sec:simulation_setup}
This section gives the settings with which the optimisation problems of Section~\ref{sec:methodology} are evaluated: the scenario and constraint values, the simulation procedure and its computational cost, and an example of what a single simulation produces.

\begin{figure}[tbp]
    \centering
    \makebox[\textwidth][c]{\includegraphics[width=\dimexpr\textwidth+16mm\relax]{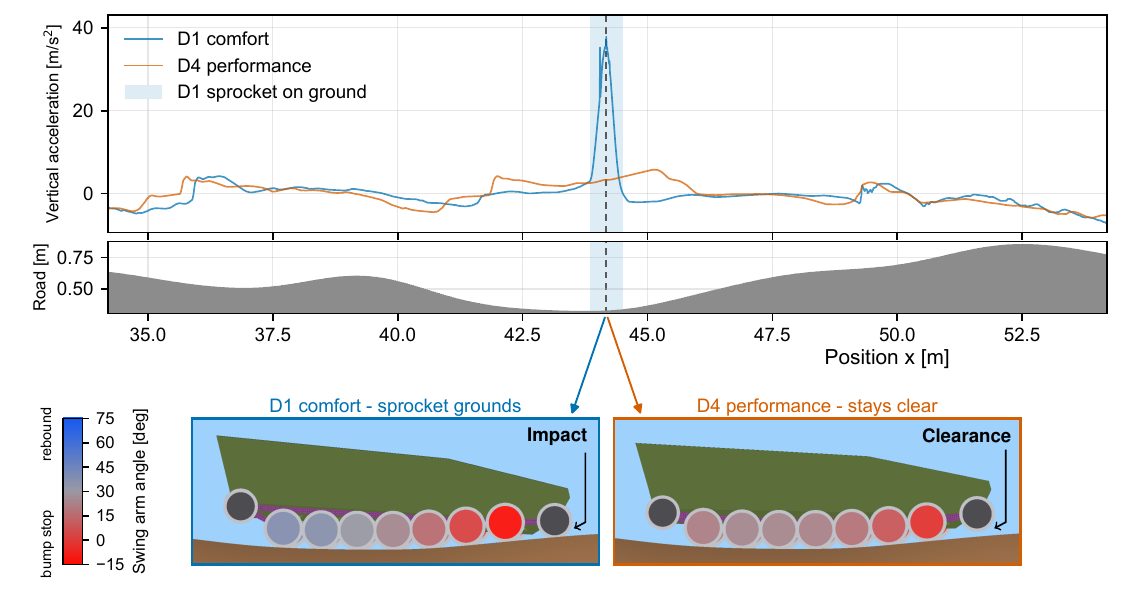}}
    \caption{Simulations on road $R_{75.6}$ at 30~km/h with the comfort design~D1 and performance design~D4. Driver vertical acceleration, road profile, and vehicle configuration at 44~s. Road wheels are coloured by swing-arm angle, from bump stop (red) to rebound (blue).}
    \label{fig:timeseries_ground_contact}
\end{figure}

\subsection{Scenario settings}
\label{sec:scenario_settings}
In formulation A, each rough road is traversed in both directions at three road-specific speeds: 40, 50 and 60~km/h on $R_{29.8}$--$R_{38.0}$, 30, 40 and 50~km/h on $R_{52.5}$ and $R_{52.7}$, 25, 30 and 35~km/h on $R_{61.8}$, and 20, 25 and 30~km/h on $R_{75.6}$. The easy roads are traversed in one direction at 30, 40, 50, 60 and 70~km/h, giving 67 scenarios in total. In formulation B, $v_{\max}$ is searched on each of the 14 rough road-directions in the range 5--70~km/h with a resolution of 1~km/h, with $\mathrm{AAP}_\text{max} = 6$~W, the criterion of TOP~01-1-014B~\citep{TOP014B}. In formulation C, the comfort objective uses the same 25 easy-road scenarios as A, and the limiting-speed objective uses the same search as in B. In all three formulations, the ride quality constraints are enforced with $\phi_\mathrm{max} = 1.0\degree$ and $c_\mathrm{min} = 0.35$~m.

\subsection{Simulation procedure and computational cost}
\label{sec:sim_procedure}
Each simulation begins on flat terrain, where the track is tensioned and the vehicle is accelerated to the target speed by controlling the drive sprocket before it enters the road. The static pitch angle and ground clearance of Eqs.~\eqref{eq:constraint-pitch}--\eqref{eq:constraint-clearance} are taken from the static equilibrium at the end of the tensioning phase.

Evaluating one design in the multi-objective formulation C requires a static simulation, 25 easy-road simulations and a limiting-speed search on each of the 14 rough road-directions, about 120 simulations in total. These run in parallel on a single 24-core compute node, up to 80 at a time, and one design takes about 10~min to evaluate. The 455 designs of this optimisation amount to about 55\,000 simulations. Most of these, about 44\,000, belong to the limiting-speed searches, of which only the 14 resulting speeds per design enter the objective. Including the time needed to fit the surrogate models and propose each new design, the optimisation ran for about 12 days.

\subsection{Example simulation}
\label{sec:example_simulation}
Figure~\ref{fig:timeseries_ground_contact} shows excerpts from two simulations on road $R_{75.6}$ at 30~km/h, comparing the comfort design~D1 and the performance design~D4 from the Pareto frontier in Section~\ref{sec:mobo_constrained}. While D1 experiences an impact at approximately 44~s, D4 maintains sufficient clearance.

\section{Results}
\label{sec:results}

This section presents the results in three steps. We first evaluate the two single-objective baselines A and B, whose shortcomings motivate the multi-objective formulation. We then present the results of the multi-objective Bayesian optimisation (MOBO) of formulation C: the Pareto frontier with four representative designs, and how the design parameters and ride quality constraints vary along it. Finally, we compare the single-objective designs with this frontier. 

\subsection{Single-objective optimisation}
\label{sec:single-objective}

First, we evaluate the two single-objective formulations A and B of Sections~\ref{sec:argminVDV} and~\ref{sec:argmaxV}.

\begin{figure}[tbp]
    \centering
    \includegraphics[width=0.75\textwidth]{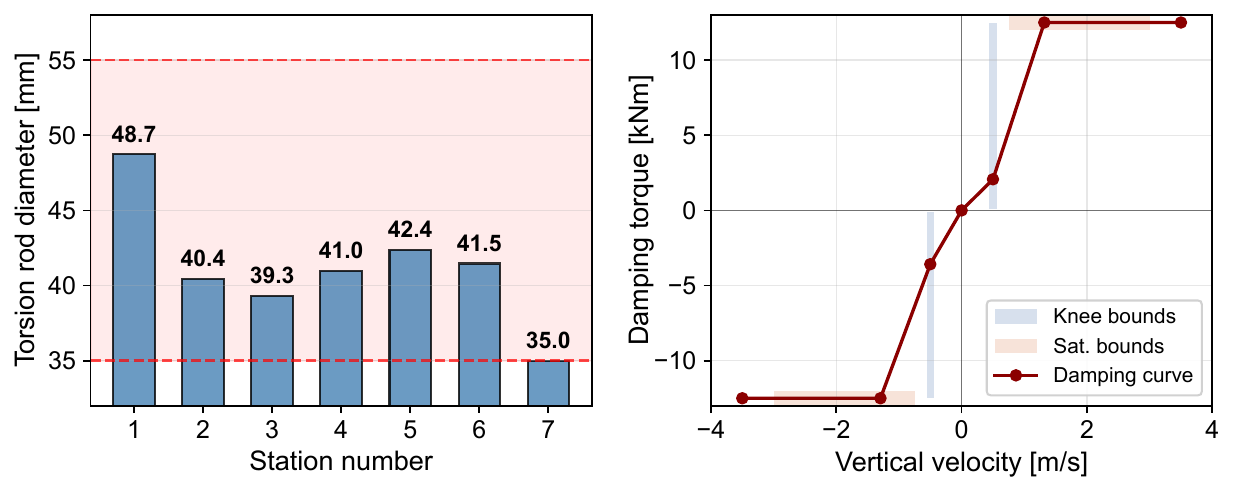}
    \caption{Best design from the pooled VDV minimisation A. Left: the seven torsion-rod diameters across the road-wheel stations (allowed range 35--55~mm). Right: the piecewise-linear damping torque versus vertical station velocity, with the fixed knee velocities at $v=\pm0.5$~m/s marked.}
    \label{fig:vdv_best_design}
\end{figure}

\paragraph*{A: Vibration minimisation}
The objective is dominated by the rough roads. For the best design, the five easy roads contribute about 3\% of the weighted VDV sum, while the two roughest roads, $R_{61.8}$ and $R_{75.6}$, contribute about half, mainly through their weights. The optimisation therefore improves the most demanding scenarios rather than typical use. 
Figure~\ref{fig:vdv_best_design} shows the best design found. The front station is stiff, stations 2--6 carry a similar, moderate stiffness, and only the rearmost station is at the lower bound, with strong damping on both sides of the curve.

\paragraph*{B: Limiting speed maximisation}

\begin{figure}[tbp]
    \centering
    \includegraphics[width=0.75\textwidth]{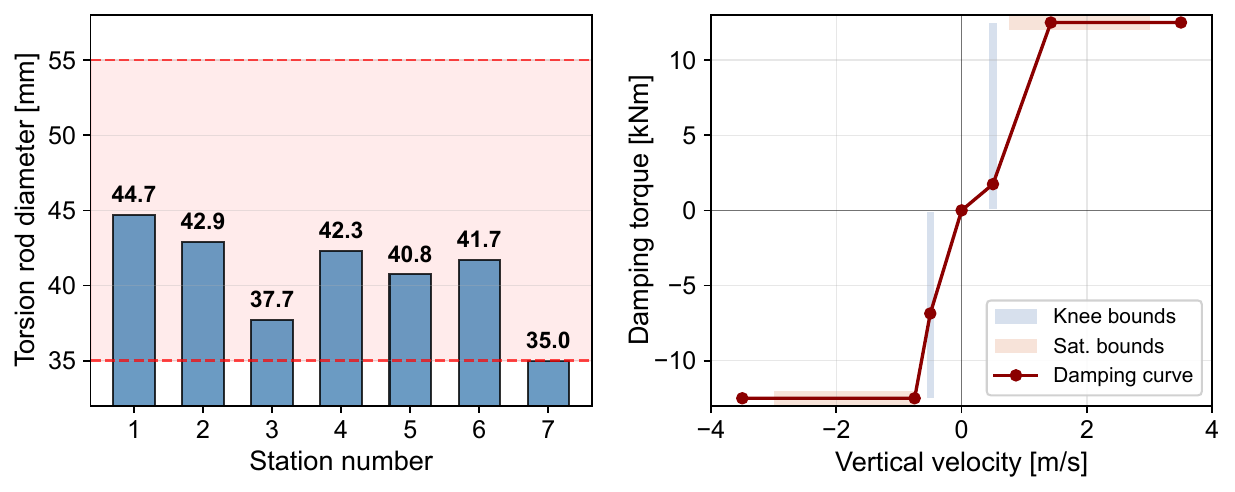}
    \caption{Best design from the limiting-speed maximisation B. Panels as in Fig.~\ref{fig:vdv_best_design}.}
    \label{fig:vmax_best_design}
\end{figure}

The limiting-speed objective of Section~\ref{sec:argmaxV} was maximised on the rough-road set. Figure~\ref{fig:vmax_best_design} shows the best design found. The layout is strikingly similar to the VDV design of Fig.~\ref{fig:vdv_best_design}: a stiff front station, similar support over stations 2--6, a soft rearmost station, and stronger compression damping.

\paragraph*{Shortcomings of the single-objective designs}
Neither solution provides comfort in routine driving. The vibration minimisation objective is dominated by the roughest scenarios, and the limiting speed maximisation objective ignores the easy roads entirely. These shortcomings motivate the multi-objective formulation that includes the pitch-comfort objective evaluated on the easy roads.

\subsection{Multi-objective optimisation}
\label{sec:mobo_constrained}

We now present the main MOBO results, rough-road speed versus pitch comfort, for the constrained two-objective problem of Eqs.~\eqref{eq:mobo-speed}--\eqref{eq:mobo-comf}.

\subsubsection*{Pareto frontier}
\begin{figure}[tbp]
    \centering
    \includegraphics[width=0.7\textwidth]{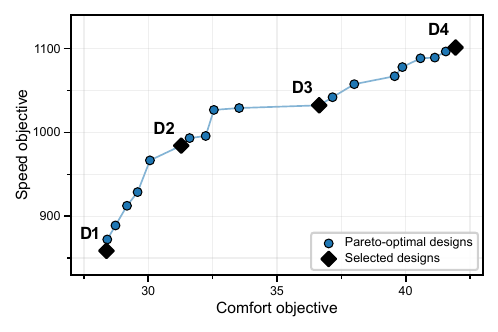}
    \caption{Pareto frontier for the main MOBO case: limiting-speed score $F_{\mathrm{speed}}$ versus pitch-VDV sum $F_{\mathrm{comf}}$. Each marker is a feasible suspension/damping design; the labelled designs D1--D4 are selected for detailed comparison.}

    \label{fig:pareto_frontier}
\end{figure}

Figure~\ref{fig:pareto_frontier} shows the resulting Pareto frontier. Each marker corresponds to a complete suspension/damping design evaluated on the rough- and easy-road scenario sets. The frontier shows a gradual transition from comfort-focused designs, with low pitch VDV, to performance-focused designs, with higher weighted limiting-speed scores. For the detailed design comparison, we select four representative Pareto designs: D1 the lowest pitch-VDV design, D2 and D3 two intermediate designs spanning the main trade-off region, and D4 the highest-performance design. These four designs are used below to compare the torsion-rod and damping settings along the frontier.

\subsubsection*{Selected designs D1--D4: suspension and damping layouts}

\begin{figure}[tbp]
    \centering
    \includegraphics[width=0.8\textwidth]{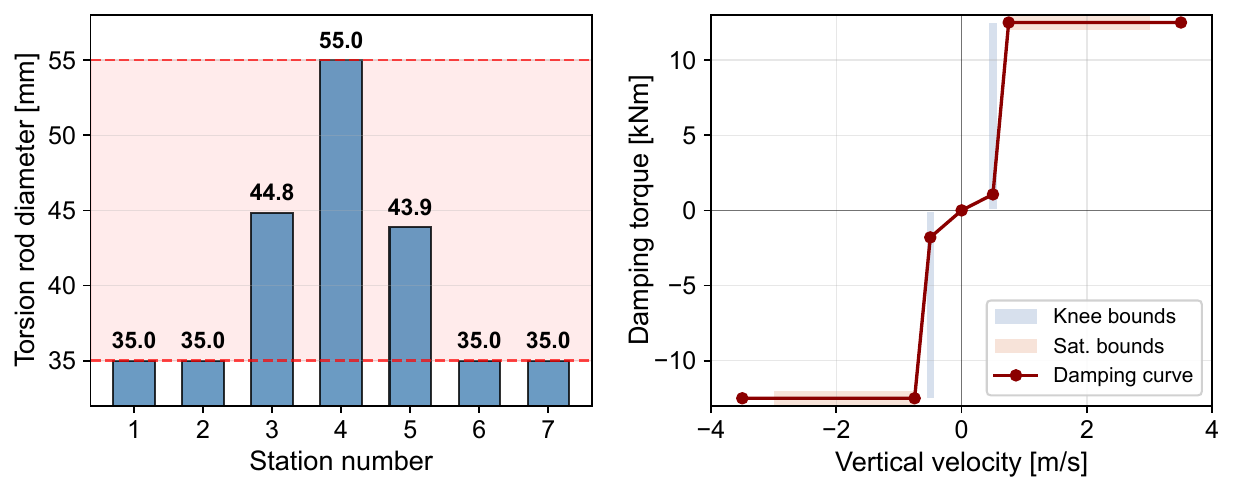}

    \vspace{0.4em}

    \includegraphics[width=0.8\textwidth]{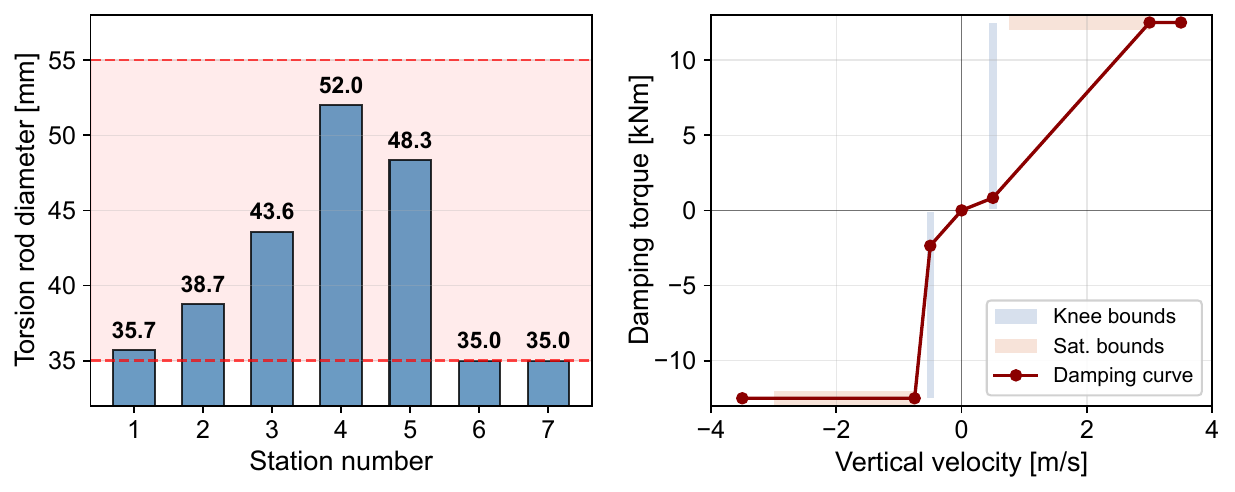}

    \vspace{0.4em}

    \includegraphics[width=0.8\textwidth]{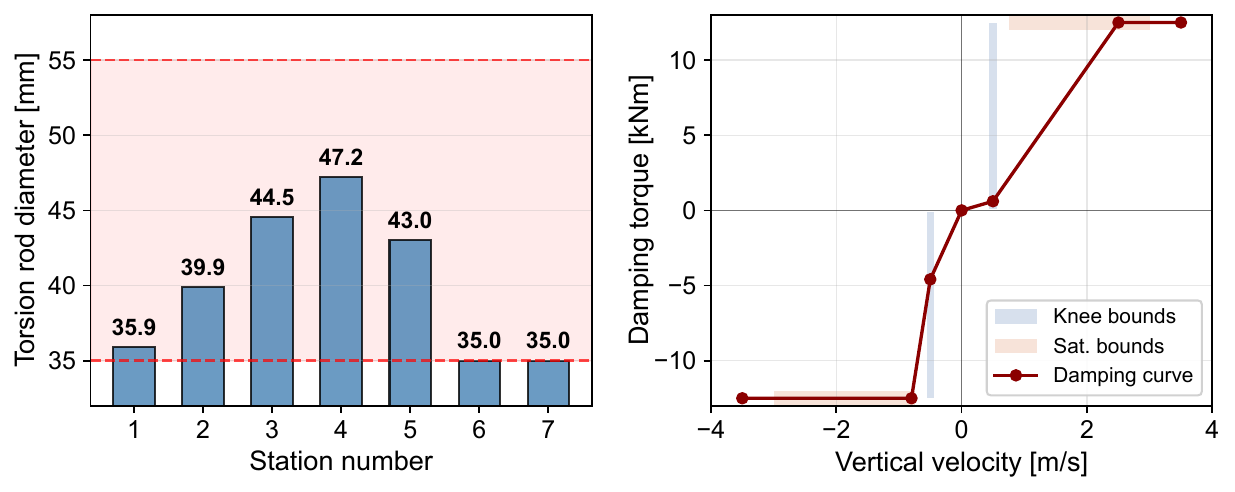}

    \vspace{0.4em}

    \includegraphics[width=0.8\textwidth]{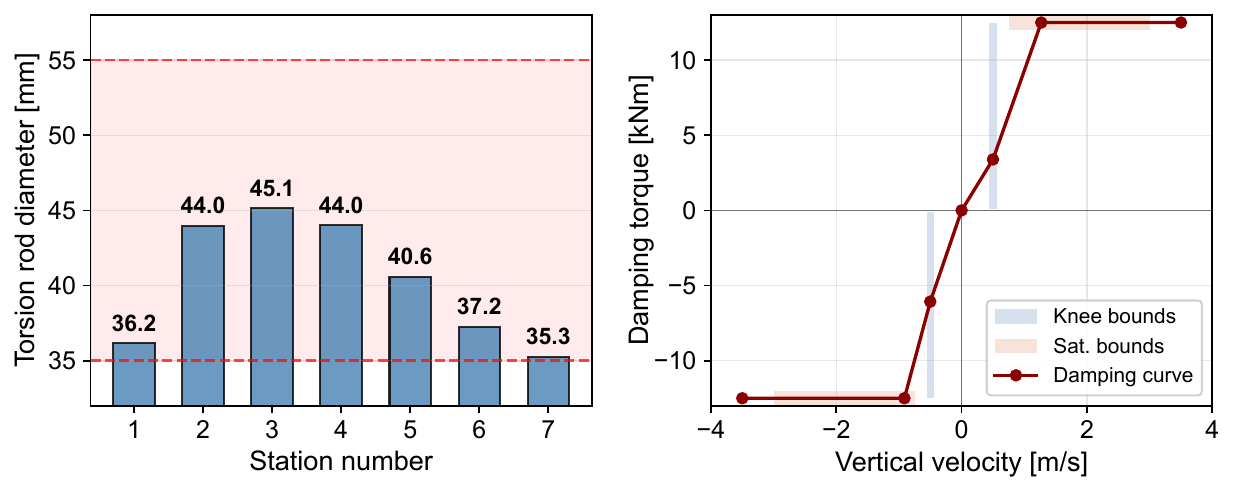}
    \caption{Torsion-rod and damping settings for the four selected Pareto designs of Fig.~\ref{fig:pareto_frontier}, from the comfort end D1 to the performance end D4. In each panel the left subplot gives the seven torsion-rod diameters across the road-wheel stations, and the right subplot gives the piecewise-linear damping torque versus vertical station velocity.}

    \label{fig:torsion_damping_1234}
\end{figure}

Figure~\ref{fig:torsion_damping_1234} compares the torsion-rod and damping settings for the four selected Pareto designs. The comfort-oriented design D1 has a strongly middle-weighted torsion-stiffness layout: rods 3--5 provide most of the suspension support, with rod 4 at its upper bound, while rods 1--2 and 6--7 are at their lower stiffness limits. This keeps suspension reaction forces small at stations far from the chassis centre of mass, reducing the pitch moments generated by easy-road excitation. Towards the performance-oriented design D4, the support spreads forward: rods 2--4 carry similar stiffness, with rod 2 raised to the level of the middle stations, while the rear stations remain soft.

The damping settings show a related trade-off through the shape of the piecewise-linear damping curve. The dampers are placed at stations 1--2 and 6--7, so damping forces there generate relatively large pitch moments about the chassis centre of mass. For the comfort designs, the damping level is kept relatively low in the part of the curve closest to zero vertical station velocity, i.e. between the fixed knee velocities at $v=\pm 0.5$~m/s. This limits damping forces during routine easy-road excitation, where the vertical station velocities are expected to remain in this low-velocity region, thereby reducing the pitch moments introduced by the dampers.

At larger vertical station velocities, outside the fixed knee points, the damping torque rises more strongly towards saturation. The comfort designs therefore retain a clearer two-regime character: low damping around zero vertical station velocity to preserve isolation, followed by stronger damping when larger swing-arm motions must be controlled. For the performance design D4, this separation becomes less pronounced, especially in compression; the curve reaches higher damping torque earlier. This is consistent with the rough-road objective, where larger terrain inputs produce higher station velocities and require more energy dissipation through the dampers. 

\subsubsection*{Trends along the Pareto frontier}

\begin{figure}[tbp]
    \centering
    \includegraphics[width=0.7\textwidth]{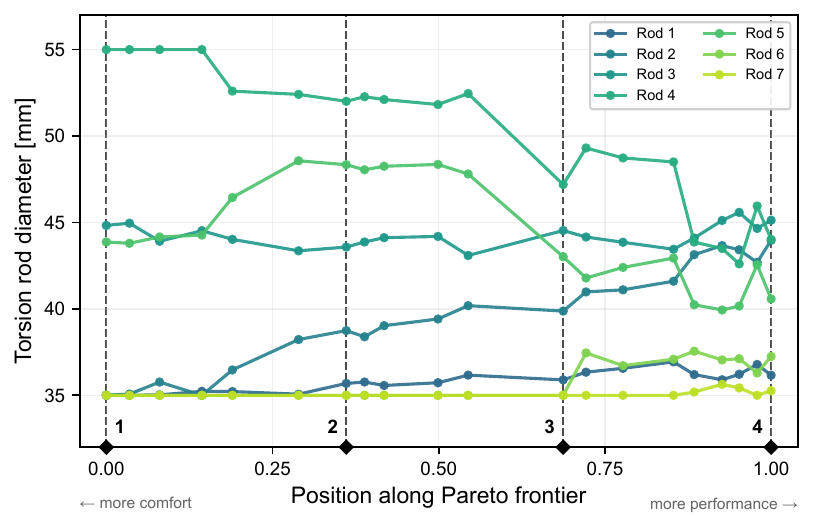}
    \caption{Torsion-rod diameters along the Pareto frontier. Each line is one of
    the seven rod diameters as a function of the position along the frontier,
    measured as normalised arc length from the comfort end (left) to the
    performance end (right).}
    \label{fig:rods_over_frontier}
\end{figure}

\begin{figure}[tbp]
    \centering
    \includegraphics[width=0.7\textwidth]{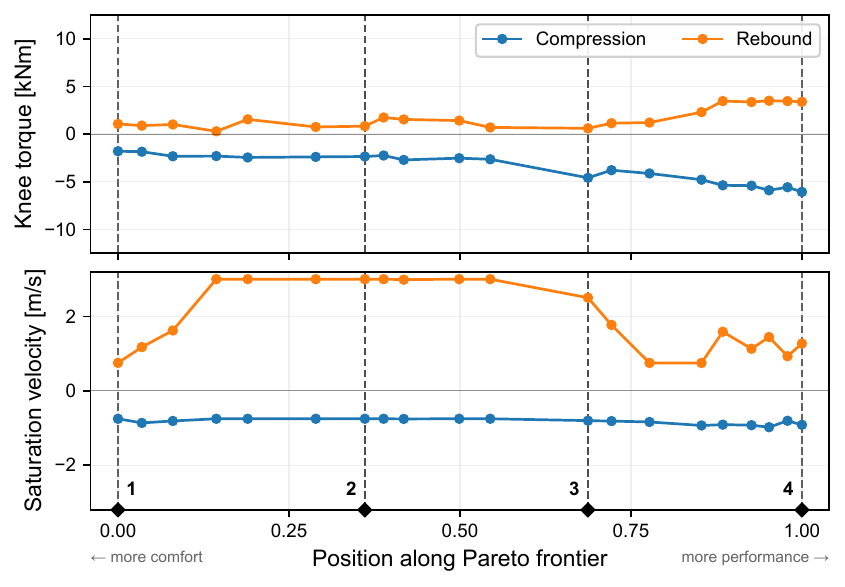}
    
    \caption{Damping parameters along the Pareto frontier, with the position
    along the frontier as in Fig.~\ref{fig:rods_over_frontier}. Top: compression
    and rebound knee torques. Bottom: compression and rebound saturation
    velocities.}
    \label{fig:damping_over_frontier}
\end{figure}

\begin{figure}[tbp]
    \centering
    \includegraphics[width=0.7\textwidth]{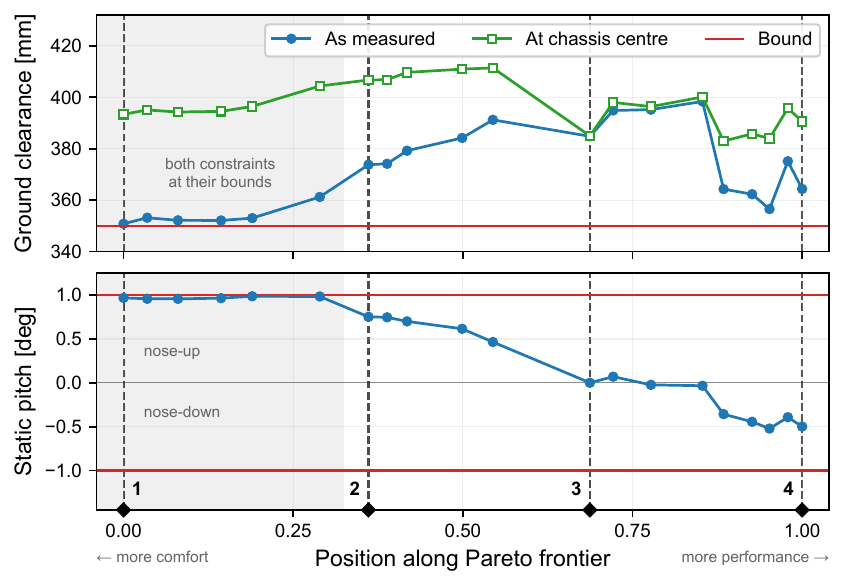}
    \caption{Ride quality constraints along the Pareto frontier, with the position along the frontier as in Fig.~\ref{fig:rods_over_frontier}. Top: ground clearance, as measured and at the chassis centre. Bottom: static pitch angle. The shaded region marks the designs with both constraints at their bounds.}
    \label{fig:frontier_constraints}
\end{figure}

Figures~\ref{fig:rods_over_frontier}, \ref{fig:damping_over_frontier} and~\ref{fig:frontier_constraints} show how the optimised parameters and the ride quality constraints vary along the Pareto frontier. The horizontal axis is the normalised arc length along the frontier, so each design is placed by how far along the frontier it lies, from the comfort end (left) to the performance end (right). These plots extend the four-design comparison by showing the full set of Pareto designs rather than only selected snapshots. 

Across the frontier, the split between designs D1 and D4 holds for the full set of Pareto designs. The comfort end reproduces the middle-weighted torsion layout of design D1, with rod 4 at its 55~mm upper bound and rods 1--2 and 6--7 at the 35~mm lower bound. Towards the performance end the stiffness distribution shifts forward: rod 2 rises steadily while rod 4 comes down off its bound, so that rods 2--4 end at a similar 44--45~mm level while the rear stations stay soft (Fig.~\ref{fig:rods_over_frontier}). The damping parameters separate the same way (Fig.~\ref{fig:damping_over_frontier}). With the knee velocities fixed, the knee torques set the slope of the damping curve near zero vertical station velocity, and both roughly triple from the comfort end to the performance end: the compression knee torque grows steadily over the performance half of the frontier, while the rebound knee torque stays low over most of the frontier and rises only close to the performance end. The transition is gradual for most parameters, with some shifting back and forth over the intermediate designs, while both ends settle into two recognisable settings. 

The compression saturation velocity stays close to the knee across the whole frontier, so beyond the knee the curve always rises steeply towards saturation (Fig.~\ref{fig:damping_over_frontier}, bottom). The rebound saturation velocity, in contrast, is not settled by the optimisation: it spans most of its range along the frontier, sitting at its upper bound over much of the intermediate region and dropping back to low values towards the performance end.

The ride quality constraints of Section~\ref{sec:constraints}, the ground clearance and pitch angle at static equilibrium, separate along the frontier as well (Fig.~\ref{fig:frontier_constraints}). For the six designs closest to the comfort end, both constraints are at their bounds. Over the intermediate designs, the clearance increases while the pitch angle decreases, passing through level at design~D3.

The two constraints are closely linked. The clearance is the minimum over the whole hull underside, so a pitched vehicle is measured at its lowest end. Measured at the chassis centre, the clearance of the Pareto designs never falls below 383~mm (Fig.~\ref{fig:frontier_constraints}, top). At the comfort end, the clearance constraint is therefore active only because of the pitch angle; at the centre, design~D1 clears the bound by 43~mm.

Near the performance end, this trend breaks: between two neighbouring designs, the clearance drops by 34~mm and the vehicle tips from level to nose-down, about equally through the tilt and a lower ride height. Because the step was unexpected, we re-evaluated both designs at fixed speeds over the full speed range, which confirmed that the second design is indeed faster. The step thus marks where the frontier reaches its performance-end setting: rods 2--4 come to a similar level, and all designs beyond it keep the nose-down attitude.

Taken together, these trends extend the picture from the selected designs. At the comfort end, soft outer stations and low damping near zero vertical station velocity keep the forces far from the chassis centre of mass small, which limits the pitch moments generated on the easy roads. These designs sit nose-up at the pitch bound, and through that tilt also at the clearance bound, with the middle rods, rod~4 at its upper bound, providing the support the soft outer stations do not. The comfort end of the frontier is thus limited by the ride quality constraints, consistent with the observation in Section~\ref{sec:constraints} that unconstrained optimisation trades static attitude for ride-metric improvements. Towards the performance end, the support moves forward and both knee torques roughly triple, dissipating the energy of the larger suspension strokes on the rough roads, while the vehicle rotates to slightly nose-down and both constraints become slack.

\subsection{Comparison with single-objective designs}
\label{sec:comparison_single}
To place the single-objective designs of Section~\ref{sec:single-objective} relative to the Pareto frontier, we re-evaluated the best designs of A and B with the scenario sets and metrics of formulation C. Since all three formulations share the same ride quality constraints, the comparison is between feasible designs. Both reach limiting-speed scores at or somewhat above the performance end of the frontier, but their pitch-VDV sums lie far above the entire frontier. Design A scores 1106.6 and design B 1160.8 against 1101 at the performance end, with pitch-VDV sums of 63.2 and 51.9 against at most 42 along the frontier. At its performance end, the multi-objective frontier thus gives up about 5\% in limiting-speed score relative to design B for about 20\% lower pitch VDV. Notably, the suspension settings at the performance end of the frontier also lie closer to the single-objective designs: the rod stiffness is spread more evenly over the front and middle stations, as in designs A and B, and the compression knee torque is close to that of design B, although stations 1 and 6 remain softer.

\section{Discussion}
\label{sec:discussion}
The three formulations, A--C, show where the problem formulation mattered most. Formulations A and B differ in their ride metric and in how speed enters the objective, yet their best designs are similar, and neither accounts for comfort in routine driving. The largest change came with formulation C, where adding the comfort objective on the easy roads moved the comfort end of the frontier to a middle-weighted rod layout with low damping near zero vertical station velocity. How far the designs could move in that direction was set by the ride quality constraints, which are active at the comfort end. Towards the performance end, the designs move closer to those of A and B.

Choices within a formulation matter as well. The weight sets how much one km/h on a road is worth, but it has little effect where the limiting speed hardly varies between designs. On the least rough roads, $v_{\max}$ reached the upper end of the search range for most designs, and on the roughest roads it rose by only a few km/h. Much of the gain along the frontier therefore came from a few roads, mainly $R_{52.5}$ and $R_{52.7}$. The number of roads per roughness level, their weights and the search range thus decide which terrain the design is tuned for. This agrees with \citet{uys2007suspension}, who found that optimal suspension settings depend on road roughness, speed and where the ride comfort is evaluated. Using several road realisations per roughness level would reduce the influence of individual road profiles. 

Although the optimisation framework identified clear trends in the suspension design, the results should be interpreted in the context of several modelling assumptions. First, the study is restricted to rigid terrain. On deformable terrain, suspension properties also influence track load distribution, sinkage, and traction, introducing additional couplings between ride quality and mobility that are not represented here. Second, the road profiles are laterally uniform and excite primarily heave and pitch. More realistic three-dimensional terrain with left-right variations, isolated obstacles, and steering manoeuvres would introduce roll dynamics and may alter the optimal suspension characteristics.

The study also deliberately focuses on ride quality rather than overall vehicle performance. In practice, suspension design must balance additional objectives such as durability, structural loads, handling, mobility on soft terrain, and energy efficiency. Extending the optimisation framework to include these competing objectives represents a natural direction for future work.

\section{Conclusion}
This paper investigated how the formulation of a simulation-based optimisation problem influences the resulting suspension design of a tracked vehicle. Three formulations were compared using the same multibody model and Bayesian optimisation framework. Minimising the driver's vertical VDV and maximising the limiting speed on rough roads led to similar designs, neither of which accounted for comfort in routine driving. Adding a pitch-comfort objective on easy roads gave a Pareto frontier ranging from a middle-weighted, lightly damped comfort design to a forward-supported, more strongly damped performance design.

The results demonstrate that optimisation problem formulation is a key component of simulation-based suspension design rather than merely part of the experimental setup. Defining objectives and evaluation scenarios that reflect the intended operational requirements is therefore essential for obtaining meaningful engineering solutions. The resulting Pareto frontier provides not only a set of optimal designs, but also a map of how the suspension design should change as the priority shifts between ride comfort and rough-terrain performance.




\section*{Declarations}

\paragraph*{Conflicts of interest}
All authors declare that they have no conflicts of interest.

\paragraph*{Use of AI technology}
The authors acknowledge the use of Claude Opus 4 and Claude Opus 5 to assist with writing and debugging the simulation, analysis and plotting code, and to improve the language and readability of the manuscript. All AI-assisted content was verified and revised by the authors to ensure authenticity, accuracy and clarity of meaning, and the authors take full responsibility for the content of the article.





\bibliography{ref}


\end{document}